\documentclass[12pt]{article}
\title{Partial and complete linearization of PDEs  \\
           based on conservation laws}
\author{Thomas Wolf \\ Department of Mathematics,
Brock University\\ 500 Glenridge Avenue, St.Catharines, 
Ontario, Canada L2S 3A1\\
email: TWolf@brocku.ca}

\begin{document}
\maketitle
\begin{abstract}
  A method based on infinite parameter conservation laws is described
  to factor linear differential operators out of nonlinear partial
  differential equations (PDEs) or out of differential consequences of
  nonlinear PDEs. This includes a complete linearization to an
  equivalent linear PDE (-system) if that is possible.  Infinite
  parameter conservation laws can be computed, for example, with the
  computer algebra package {\sc ConLaw}.
\end{abstract}

\begin{center}
\end{center}

\section{Introduction} 
With the availability of computer algebra programs for the automatic
computation of all conservation laws up to a given differential order
of the integrating factors 
(as described in \cite{WBM}, \cite{CL4})
conservation laws have been found that involve arbitrary functions,
i.e.\ infinitely many parameters.
In this paper we show how based on such conservation laws 
a linear differential operator can be factored out of a combination of
the original nonlinear partial differential equations
(PDEs) and their differential consequences. Possible outcomes
include 
\begin{itemize}
\item a complete linearization into an equivalent linear system,
\item a partial linearization in the sense that a linear differential
  operator is factored out, splitting the problem into a linear one
  plus a non-linear problem of lower order and often fewer independent
  variables (e.g.\ ordinary differential equations (ODEs)),
\item the derivation of at least one linear equation from a nonlinear
  system (with the possibility of deriving further linear equations
  for the new mixed linear-nonlinear system).
\end{itemize}
An advantage of the procedure to be presented is that
conservation laws need not be given explicitly in terms of 
the arbitrary functions. It is enough to
have the conservation law determining conditions solved up
to the solution of a system of consistent and necessarily linear PDEs which 
have arbitrary functions in their general solution.

The content of the paper is as follows. After comments are made on 
the computation of conservation laws in section \ref{claf}, 
the four computational steps of factoring out linear differential operators
are illustrated using the Liouville equation in section \ref{method}. 
Sufficient conditions for complete or partial linearizations
are listed in section \ref{clid}, followed by a discussion of computational
aspects in section \ref{CompAsp}.
A generalization involving the introduction of potentials 
in terms of which a linearization becomes possible 
is explained in section \ref{secex}.
In later sections \ref{inhomsec}, \ref{triagsection},
an illustration is given of how the method works 
when nonlinear equations linearize to
{\em inhomogeneous} equations or to {\em triangular linear} systems. 
Further examples where a complete or at least 
a partial linearization is possible are given in the appendix.

In this contribution we concentrate on computational aspects of the
method and give examples for all of the above scenarios. 
An extension of the method discussing complete and partial linearizability 
through point and contact transformations 
will appear in a future publication \cite{AWB},
with numerous new examples
and a comparison with other linearization methods found in the literature. 

\section{Notation} 
We follow the notation in \cite{PO2}
and denote the original nonlinear partial differential 
equations as $0 = \Delta_{\alpha}$, the dependent variables
by $u^{\beta}, \;\; \alpha, \beta=1\ldots q$ and the independent variables 
by $x^i, \;\;i=1\ldots p$. In examples dealing with functions
$u=u(x,t)$ or $u=u(x,y)$, partial derivatives
are written as subscripts like $u_{xy}=\partial^2u/(\partial x\partial y)$. 
If a formula already contains subscripts then $\partial_i$ will be
used for $\partial/\partial x^i$. The multi indices $_J, _K$ denote multiple
partial derivatives like $u^\alpha_J$ which in our notation 
include $u^\alpha$. With $\#J$ we denote the differential order,
i.e.\ number of partial derivatives represented by $_J$.
Total derivatives with respect to $x^i$ will
be denoted as $D_i$. We apply the convention that
summation is performed over terms that involve two identical 
indices, one subscript and one superscript. For example, 
the divergence of a vector field $P^i$ would be denoted as 
$D_i P^i \ (\equiv \sum_i D_i P^i)$.
The procedure to be presented repeatedly uses adjoint differential operators
as follows. For given functions $f^A(x^i),\ A=1..r$, let linear
differential expressions $H_k$ be defined as
\[ H_k= a^J_{kA} \partial_J f^A , \ \ k=1,\ldots,s\ , \]
with coefficients $a^J_{kA}=a^J_{kA}(x^i)$ 
and summation over $A$ and the multi index $J$.
The corresponding adjoint operators $H^{\;*}_{Ak}$ are 
computed for arbitrary functions $G^k(x^i)$ by repeatedly reversing 
the product rule of differentiation for the sum $G^k H_k$ to get 
\begin{equation}
 G^k H_k = f^A H_{Ak}^{\;*} G^k + D_i \bar{P}^i  \label{not1}
\end{equation}
where
\begin{equation}
 H_{Ak}^{\;*} G^k = (-1)^{\#J}\partial_J\left(a^J_{kA}G^k\right) .  \label{not2}
\end{equation}
and $\bar{P}^i$ are expressions resulting from integration by parts
with respect to $\partial_J$ in this computation.

\section{Conservation laws with arbitrary functions} \label{claf} 
Conservation laws can be formulated in
different ways (see \cite{CL4} for four different approaches to
compute conservation laws). The form to be used in this paper is
\begin{equation}
 D_i P^i = Q^\alpha \Delta_\alpha         \label{cl2}
\end{equation}
where the components $P^i$ of the conserved current {\bf} 
and the so-called characteristic functions $Q^\alpha$ are differential
expressions involving $x^i, u^\alpha_J$. Other forms of conservation
laws can easily be transformed into (\ref{cl2}).
One approach to find conservation laws for a given system of
differential equations $0=\Delta_\alpha$ is to 
specify a maximum differential order $m$ of derivatives $u^\alpha_J$ on
which $P^i, Q^\alpha$ may depend and then to solve condition
(\ref{cl2}) identically in $x^i, u^\alpha_J$ for the unknown functions
$P^i, Q^\alpha$. Due to the chain rule of differentiation in (\ref{cl2}) 
the total derivatives $D_i$ introduce extra 
derivatives $u^\alpha_K$ with $\#K=m+1>m$, i.e.\ derivatives not occuring as 
variables in $P^i, Q^\alpha$. Splitting with
respect to these $u^\alpha_K$ results in an {\em overdetermined} and {\em linear}
system of PDEs for $P^i, Q^\alpha$.\footnote{Note that
  regarding (\ref{cl2}) as an algebraic system for unknowns $Q^\alpha$
  implies division through $\Delta_\alpha$ and
  does therefore not produce $Q^\alpha$ which are regular for solutions
  $u^\alpha$ of the original system $\Delta_\alpha=0$. For details
  regarding the ansatz for $Q^\alpha$ see \cite{CL4}.}

What is important in the context of this paper is that a differential
Gr\"{o}bner basis can be computed algorithmically and from it
the dimension of the solution space can be determined, 
i.e.\ how many arbitrary functions of how many variables
the general solution for $P^i, Q^\alpha$ depends on. In extending the
capability of a program in solving conditions (\ref{cl2}) 
by not only computing a
differential Gr\"{o}bner basis (for linear systems) but also integrating
exact PDEs and splitting PDEs with respect to only explicitly occuring
$u^\alpha_J$ (which here act as independent variables), the situation
does not change qualitatively. The result is still either the explicit
general solution or a linear system of unsolved PDEs 
\begin{equation} 
0 = C_k(x^i, u^\alpha_J, f^A) , \ \ \ \ k=1,\ldots,r\ ,      \label{remcond}
\end{equation}
for some functions $f^A(x^j, u^\beta_J)$ where this system is 
a differential Gr\"{o}bner basis 
and allows one to determine algorithmically the size of the solution 
space. The functions $f^A$ are either the $P^i, Q^\alpha$ themselves
or are functions arising when integrating the
conservation law conditions (\ref{cl2}).

If the conservation law condition (\ref{cl2}) is solved, i.e.\
$P^i, Q^\alpha$ are determined in terms of $x^i, u^\alpha_J, f^A_K$
possibly up to the solution of remaining conditions (\ref{remcond})
then it is no problem to use a simple division algorithm to determine
coefficients $L^k$ satisfying
\begin{equation}
Q^\alpha \Delta_\alpha = D_i P^i + L^k C_k    \label{id1}
\end{equation}
identically in $x^i,u^\alpha_J, f^A_J$. 
The coefficients $L^k$ are necessarily free of $f^A_J$ because 
(\ref{cl2}) is linear and homogeneous in $Q^\alpha, P^i$ 
and this property is preserved in solving these conditions, 
so $C_k$ are  linear and homogeneous in $f^A_J$ as well
and $L^k$ must therefore be free of $f^A_J$.
We will call relation (\ref{id1}) a `conservation law identity'
because it is satisfied identically in all $x^i, u^\alpha_J$ and
$f^A_J$.  

\section{The Procedure}  \label{method} 
The individual steps of our method are
shown in detail to demonstrate that all steps are algorithmic and can
be performed by computer. 
The {\sc Reduce} package {\sc ConLaw} has the algorithm implemented
and performs it whenever a conservation law computation results in a
solution involving arbitrary functions possibly up to the solution of a linear
system (\ref{remcond}). \vspace{6pt}

{\bf Input} \\
Input to the procedure is the conservation law identity (\ref{id1}) 
\begin{equation}
Q^\alpha \Delta_\alpha = D_i P^i + L^k C_k   
\end{equation}
including expressions for all its constituents 
$Q^\alpha, P^i, L^k, C_k$ in terms of $x^i, u^\alpha, f^A$.

To start the procedure the functions $f^A$ have to depend only on
the variables $x^i$. If they depend on $u^\alpha_J$ then a
linearizatin will necessarily involve a change of variables. This case
is treated in \cite{AWB}.

{\bf Step 1.} \\
If all functions $f^A$ depend exactly on all $p$
independent variables $x^i$ then proceed to step 2. Step 1 is
concerned with the case that not all $f^A=f^A(x^i)$ depend on
all $x^i$. To add the dependence of, say $f^B$ on $x^j$ one
has to 
\begin{itemize} 
\item compute 
\[ Z:=\left.(Q^\alpha \Delta_\alpha - D_i P^i - L^k C_k)
      \right|_{f^B(x^i)\rightarrow f^B(x^i,x^j)} \]
which vanishes modulo $0=\partial_j f^B$ and therefore must
have the form 
\[ Z= M^J \partial_J\left(f^B\right) \]
with suitable coefficients $M^J$ and summation over the multi index $J$,
\item compute the adjoint $Z^*_B$ as in (\ref{not1}),(\ref{not2})
to bring $Z$ into the form
\begin{equation}
 Z=D_i \bar{P}^i + \partial_jf^B Z^*_B , \label{adj1}
\end{equation}
\item rename $P^i + \bar{P}^i \rightarrow P^i$ and adds a new
  condition $C_{r+1}=\partial_jf^B$ and multiplier
  $L^{r+1}=Z^*_B$ to arrive at a new version of the conservation law
  identity $Q^\alpha \Delta_\alpha = D_i P^i + L^k C_k$ where the
  function $f^B$ depends now on $x^j$.
\end{itemize}
This process is repeated until all $f^A$ depend on all $x^i$.

{\em Example 1:}
We will illustrate the steps of the procedure with an investigation of the
Liouville equation 
\begin{equation}
0 = \Delta := u_{xy} - e^u.   \label{liou}
\end{equation}
Although it is not completely linearizable, we choose this equation
because it involves computations in each of the first three steps.

For the Liouville equation a conservation law identity
involving an arbitrary function $f(x)$ is given through
\begin{equation}
(f_x+fu_x) \Delta = D_x(-f e^u) + D_y(f_xu_x + fu_x^2/2), \label{liou1}
\end{equation}
i.e.\ $Q=f_x+fu_x,\;\,P^x=-f e^u,\;\,P^y=f_xu_x + fu_x^2/2,\;\,C_k=0$.
Adding a $y$-dependence to $f$ requires to add to the right hand side
of our identity (\ref{liou1}) the terms
\[Z = - f_{xy}u_x - f_yu_x^{\;\,2}/2\]
which in adjoint form (\ref{adj1}) read 
\[Z = D_x(-f_yu_x) + (u_{xx}-u_x^{\;\,2}/2)f_y \]
giving the new conservation law identity
\begin{equation}
(f_x+fu_x) \Delta = D_x(-f e^u-f_yu_x) + D_y(f_xu_x + fu_x^2/2) 
+ (u_{xx}-u_x^{\;\,2}/2)f_y . \label{liou2}
\end{equation}

{\bf Step 2.} \\
As the $Q^\alpha$ are linear homogeneous differential expressions for the $f^A$
one can compute adjoint operators $Q^{\alpha *}_A$
as in (\ref{not1}),(\ref{not2}) by expressing 
\[ Q^\alpha \Delta_\alpha = f^A Q^{\alpha *}_A \Delta_\alpha 
                            + D_i\bar{P}^i. \]
After renaming $P^i - \bar{P}^i \rightarrow P^i$
the conservation law identity takes the new form
\begin{equation}
 f^A Q^{\alpha *}_A \Delta_\alpha = D_i P^i + L^k C_k. \label{id2}
\end{equation}
In the case of the Liouville equation we partially integrate
\[ (f_x+fu_x) \Delta = f (u_x -D_x)\Delta + D_x(f\Delta) \]
and get the conservation law identity 
\begin{eqnarray}
 f (u_x -D_x)\Delta &=& D_x(-f e^u-f_yu_x-f\Delta) + D_y(f_xu_x + fu_x^2/2) 
                      + (u_{xx}-u_x^{\;\,2}/2)f_y \nonumber \\
                    &=& D_x(-f_yu_x-fu_{xy}) + D_y(f_xu_x + fu_x^2/2) 
                      + (u_{xx}-u_x^{\;\,2}/2)f_y . \label{liou3} 
\end{eqnarray}

{\bf Step 3.} \\
Because the $C_k$ are linear homogeneous differential expressions in the $f^A$ 
we can compute the adjoint form of $L^kC_k$ as in 
(\ref{not1}),(\ref{not2}) by expressing 
\[ L^kC_k = f^AC^{\;*}_{Ak}L^k + D_i\bar{P}^i .\]
After renaming $P^i + \bar{P}^i \rightarrow P^i$
the conservation law identity takes the new form
\begin{equation}
 f^A Q^{\alpha *}_A \Delta_\alpha = D_i P^i + f^AC^{\;*}_{Ak}L^k. \label{id3}
\end{equation}
In our example partial integration gives
\[(u_{xx}-u_x^{\;\,2}/2)f_y = 
  D_y((u_{xx}-u_x^{\;\,2}/2)f) - f(u_{xx}-u_x^{\;\,2}/2)_y\]
and substituted into (\ref{liou3}) the new conservation law identity
\begin{eqnarray}
 f (u_x -D_x)\Delta &=& D_x(-f_yu_x-fu_{xy}) 
                      + D_y\left(f_xu_x + fu_x^2/2 + f(u_{xx}-u_x^{\;\,2}/2) \right) 
                      - f(u_{xx}-u_x^{\;\,2}/2)_y  \nonumber \\ 
                   &=& - f(u_{xx}-u_x^{\;\,2}/2)_y \label{liou4} 
\end{eqnarray}
after simplification.

{\bf Step 4.} \\
This step does not involve any computation, it merely completes the
constructive proof how linearizations are achieved.

By bringing $f^A C_{Ak}^{\;*} L^k$ to the left hand side of the
conservation law identity (\ref{id3}) we get
\begin{equation}
f^A \left(Q^{\alpha *}_A \Delta_\alpha - C_{Ak}^{\;*} L^k\right) = 
D_i P^i     \label{step4}
\end{equation}
which still is an identity for arbitrary functions $u^\alpha, f^A$.
Applying the Euler operator with respect to $f^A$ 
(for its definition see e.g.\ \cite{PO2},\cite{AWB})
to the left hand side of  (\ref{step4}) gives the coefficient of $f^A$
and on the right hand side gives zero as it is a 
divergence,\footnote{To prove this statement without Euler operator
we could choose the $f^A$ to be zero outside some region
$R$ such that an integral over a volume with boundary outside $R$
will vanish using Gauss law on the right hand side of identity
(\ref{step4}) as $P^i$ are linear homogeneous in the $f^A$.
Because the $f^A$ are arbitrary inside $R$ the coefficients of the
$f^A$ on the left hand side must vanish identically.}
i.e.\ we get
\begin{equation}
Q^{\alpha *}_A \Delta_\alpha = C_{Ak}^{\;*} L^k \;\;\;\;\;\;\;\;\;\;
\mbox{identically in}\;\; u^\alpha \;\;\mbox{for all} \;\; A. \label{lin}
\end{equation}
The vanishing of $D_i P^i$ on the right hand side of (\ref{liou4}) 
was therefore not accidental.

For the Liouville equation the identity (\ref{lin}) takes the form
\begin{eqnarray}
(u_x  - D_x) \Delta  & = & - D_y L\;\,=\;\,0 \;\;\;\;\;\;\;\;\;\;\;\;\;\;\;\;
\mbox{with} \label{liouD} \\
L & = & u_{xx} - u_x^{\;\,2}/2.  \label{liouL}
\end{eqnarray}
Integrating at first (\ref{liouD}) to $L=L(x)$ leaves the Riccati ODE
\begin{equation}
 u_{xx} - u_x^2/2 = L(x)       \label{linliou}
\end{equation}
for $u_x$ to be solved, for example, through a linearizing
transformation $u(x,y)= -2\log(v(x,y))$.

{\bf Output} \\
The result of the procedure are expressions $Q^{\alpha *}_A,
C_{Ak}^{\;*}$ and $L^k$. The relation
\begin{equation}
C_{Ak}^{\;*} L^k = 0 \label{linend1}
\end{equation}
is a necessary condition which can be solved by first regarding
$L^k$ as dependent variables and then solving 
\begin{equation}
L^k=L^k(u^\alpha_J)  \label{linend2}
\end{equation}
for $u^\alpha=u^\alpha(x^i)$.
The system (\ref{linend1}), (\ref{linend2}) is 
a sufficient condition for the original system $\Delta_\alpha=0$ if 
$Q^{\alpha *}_A$ is an invertible algebraic operator 
and it is a complete linearization if (\ref{linend2}) is purely
algebraic in $u^\alpha$.

\section{Scope of the procedure} \label{clid} 
The degree to which the original system
$\Delta_\alpha=0$ can be linearized depends on properties of the
conservation law identity that has been computed: 
the number of functions $f^A$ and the
number of variables each $f^A$ depends on, the differential
order of derivatives of $f^A$ with respect to $x^i, u^\alpha_J$ in 
$C_k$ and in $Q^\alpha$.  Some properties, like the size of the
solution space of remaining conditions (\ref{remcond}) are essentially
independent of the extent to which these conditions are solved. Other
criteria, like the number of functions $f^A$ and the number of their
arguments does depend on the extent to which conditions (\ref{remcond})
were solved. The strength of the procedure to be presented is to be
able to handle a wide range of situations.

The following is a list of five scenarios, sorted from most special,
fully algorithmic (and most beneficial) to most general, not strictly
algorithmic (and less beneficial).  We refer to the computational
steps described in section \ref{method} as `the procedure'.

\itemsep 0pt 
\begin{itemize}
\item If the following criteria are met: \begin{enumerate} 
  \item the size of the solution space of $0=\Delta_\alpha$ is equal to
    the size of the solution space of $0=C_k$,
  \item the conditions $0=C_k$ involve $q$ functions $f^A$, (equal to the
    number of functions $u^\alpha$ in $\Delta_\alpha$) and all $f^A$
    depend on $p$ variables (equal to the number of variables $u^\alpha$
    depend on), 
  \item the functions $Q^\alpha$ expressed in terms of $f^A$ involve
    $f^A$ only algebraically, i.e.\ no derivatives of $f^A$, 
  \item functions $f^A$ do not depend on jet variables $u^\alpha_J$,
    i.e.\ $f^A=f^A(x^i)$,  
  \end{enumerate}
  then the procedure will algorithmically provide a linearization of the system
  $\Delta_\alpha=0$.

  {\em Example 2:} The Burgers equation in the form
  \begin{equation}
  0 = \Delta_1 := u_t - u_{xx} - uu_x       \label{burg1}
  \end{equation}
  for a function $u(x,t)$ can not be linearized but in the potential form
  \begin{equation}
  0 = \Delta_2 := v_t - v_{xx} - v_x^{\;\;2}/2       \label{burg2}
  \end{equation}
  for $v(x,t)$ 
  a conservation law identity involving a function $f(x,t)$ is given through
  \begin{equation}
  f e^{v/2} \Delta = D_t \left(f2e^{v/2}\right) +
                     D_x \left(f_x2e^{v/2} - fe^{v/2}v_x\right)
                     +2e^{v/2}(-f_t-f_{xx})
  \end{equation}
  and the related linearization is
  \begin{eqnarray*}
  L & = & 2e^{v/2} \\
  e^{v/2}\Delta & = & L_t - L_{xx}\;\;=\;\;0.
  \end{eqnarray*}
  A proof that every non-linear PDE (-system) that is linearizable
  through point or contact transformations can be linearized this way 
  will be given in \cite{AWB}.
\item If criteria 1,2,3 are satisfied but not 4 then
  a linearization is possible but at the price of a change of
  variables, which will be a point or contact transformation
  if it is invertible or otherwise it will be a non-invertible transformation
  depending on derivatives of $u^\alpha$. 
  Furthermore, in all such cases the transformation can be derived explicitly
  from the conservation law identity as will be shown in \cite{AWB}. 
\item If criterion 3 is not satisfied then the partially or completely
  linearized equations may only be a necessary but not a sufficient
  condition for $\Delta_\alpha=0$.
\item If criterion 1 is satisfied but not 2 then  \begin{itemize} 
  \item if functions $f^A$ of fewer than $p$ variables occur then one
    can add extra variable dependencies through step 1 of the procedure,
  \item if more than $q$ functions $f^A$ occur in $0=C_k$ or functions
    $f^A$ of more than $p$ variables occur then one has to integrate
    more of the conditions $0=C_k$ in order to be able to linearize the
    original system completely (a full treatment of this case will be given in 
    \cite{AWB}).
  \end{itemize}
\item If criterion 1 is not satisfied but the solution space of $C_k$
  involves at least one arbitrary function of one argument then the
  method will result in a differential expression for $u^\alpha_J$
  which vanishes modulo $0=\Delta_\alpha$ and factorizes into a linear
  differential operator acting on a non-linear differential expression.
  Typically this leads to a PDE for $u^\alpha$ which is lower in
  differential order than $\Delta_\alpha$ for one of the $x^i$. In
  example 1 in section \ref{method} and examples 8,9,10 in the 
  appendix an equation in one less variable results, i.e.\ an ODE.
\end{itemize}
The algorithmic beauty of the procedure is that the above wide range
of situations are covered by one and the same 4-step algorithm.

The case that a non-local linearization exists in which the $L^k$
depend on integrals of $u^\alpha$ is not covered directly as the
computer algebra package {\sc ConLaw} does not compute non-local 
conservation laws. 
On the other hand single conservation laws (without parametric
functions) can be used to introduce potentials such
that the original system re-formulated in these potentials is
linearizable. This approach has been successful in all 6 linearizable
evolutionary systems found in \cite{SW}. Examples given in this paper
are the system (\ref{pot1}), (\ref{pot2}) in the section \ref{secex},
the system (\ref{tri1}), (\ref{tri2}) in section \ref{triagsection}
and the system (\ref{vol41}), (\ref{vol42}) in the appendix.


\section{Computational Aspects} \label{CompAsp} 
Given a nonlinear PDE system $0=\Delta_\alpha$, what are possible
computational hurdles to be overcome in oder to find a linearization?
The method described in section \ref{method} is algorithmic and does
not pose a problem. The formulation 
of conservation law conditions (\ref{cl2}) and their analysis through
computing a differential Gr\"{o}bner basis $0=C_k$ is algorithmic as
well and could only become difficult because of a growing size of 
equations. 

{\em A first computational challenge lies in the fact that for
  linearizable systems $0=\Delta_\alpha$ the conservation law
  conditions (\ref{cl2}) have a general solution involving arbitrary
  functions.}  It is well known that systems of equations with a large
  solution space are much harder to solve than systems with only few
  solutions or no solutions.  To incorporate many solutions, algebraic
  Gr\"{o}bner bases for algebraic systems have to be of high degree
  and differential Gr\"{o}bner bases for differential systems have to
  be of sufficiently high differential order. As a consequence, the
  length of expressions encountered during the Gr\"{o}bner basis
  computation is more likely to explode and exceed available resources.

{\em The second challenge is to integrate a Gr\"{o}bner basis $0=C_k$
sufficiently often to meet criterion 2 in section \ref{clid}.}
Because the general solution of the conservation law conditions
involves arbitrary functions, any integrations to be done can only be
integrations of PDEs, not of ODEs.

The package {\sc Crack} that is used to compute the examples in this
paper differs from similar other programs (as listed in \cite{WHER})
in that it has a number of modules addressing the above problems. For
example, the growth of expressions is lowered by a module for reducing
the length of equations by replacing them through a suitable linear
combination of equations as described in \cite{SR}.  Integrations are
handled by a module that integrates exact PDEs, that is able to
introduce potentials to integrate certain generalizations of exact
PDEs and that determines monomial integrating factors to achieve
integration (see \cite{SIE}).  A relatively new module applies
syzygies that result as a by-product of a differential Gr\"{o}bner
basis computation. This module allows to perform integrations more
efficiently and to avoid a temporary explosion of the number of
functions of integration generated in the process (see \cite{SY}). A
module that integrates underdetermined linear ODEs with non-constant
coefficients is often useful in the last stages of the computation.  A
description of the algorithm and its implementation is in preparation.

\section{An example requiring the introduction of a
         potential}\label{secex} 
The following example demonstrates that a linearization of a
non-linear equation or system may only be possible 
if it is reformulated in terms of potentials which in turn might 
be found by studying conservation laws. 

{\em Example 3:} The system 
\begin{eqnarray}
 0 \;\,= \;\,\Delta_1 & := & u_t - u_{xx} - 2 v u u_x  - 2 ( a + u^2) v_x 
                     - v^2 u^3 - b u^3 - a u v^2 - c u  \label{pot1} \\
 0 \;\,= \;\,\Delta_2 & := & v_t + v_{xx} + 2 u v v_x  + 2 ( b + v^2) u_x 
                     + u^2 v^3 + a v^3 + b v u^2 + c v  \label{pot2}
\end{eqnarray}
with $u=u(x,t),\; v=v(x,t)$ and constants $a,b,c$
results as one of 15 cases of a class of generalized non-linear
Schr\"{o}dinger equations \cite{SW}. This system itself
does not have conservation laws involving arbitrary functions 
but it has the zeroth order conservation law
\begin{equation}
v \Delta_1 + u \Delta_2 = D_t (uv) + D_x (v_xu - u_xv + b u^2 - a v^2)
\nonumber
\end{equation}
which motivates the introduction of a function $w(x,t)$ through
\begin{eqnarray}
   w_x & = & uv,                                  \label{ex10c}   \\
 - w_t & = & v_xu - u_xv + b u^2 - a v^2  .     \nonumber 
\end{eqnarray}
The remaining system to be solved for $r:=u/v$ and $w$ simplifies
if we substitute 
\begin{equation}
w = \frac{1}{2} \log z                          \label{ex10wz}
\end{equation}
with $z = z(x,t)$. This substitution is not essential for the
following but it reduces the size of the resulting system for $r(x,t), z(x,t)$
and eases memory requirements in the computation of conservation
laws of the resulting system $\Delta_3, \Delta_4$:
\begin{eqnarray}
0 \; = \; \Delta_3 & := & 2 r r_t z_x^2 + r_x^2 z_x^2 + 2 a r^2 r_x z_x^2
                          - 2 b r_x z_x^2 + 2 r^2 z_x z_{xxx}  \nonumber \\
                   &    & - r^2z_{xx}^2 + 2ar^3z_xz_{xx} + 2brz_xz_{xx}
                          + 4cr^2z_x^2 \vspace{6pt}  \nonumber \\ 
0 \; = \; \Delta_4 & := & r_xz_x + rz_t - a r^2 z_x + b z_x.  \label{ex10f}
\end{eqnarray}
The program {\sc ConLaw} finds a conservation law 
with integrating factors
\begin{eqnarray*}
Q^3 & = &  r^{-5/2} z_x^{\;\,-3/2}(fr + \tilde{f})  \\
Q^4 & = & r^{-5/2} z_x^{\;\,-3/2}
          \left( - 2z_xr(f_xr-\tilde{f}_x) - r_xz_x(fr + \tilde{f})
                              + z_{xx}r(fr - \tilde{f}) \right) 
\end{eqnarray*}
involving two functions $f(x,t), \tilde{f}(x,t)$ that have 
to satisfy the conditions
\begin{eqnarray*}
0 \; = \; C_1 & := &   - f_t + f_{xx} + c f - 2 a \tilde{f}_x  \\
0 \; = \; C_2 & := & \;\;\;\tilde{f}_t  + \tilde{f}_{xx} + c \tilde{f}
                      - 2 b f_x .
\end{eqnarray*}
The conservation law identity takes the form
\begin{equation}
Q^3 \Delta_3 + Q^4 \Delta_4 = D_t P^t + D_x P^x + 
                              L^1 C_1 + L^2 C_2   \label{ex10m}
\end{equation}
with some conserved current $(P^t, P^x)$ and 
coefficients $L^1,L^2$ of $C_1,C_2$
\begin{equation}
L^1 = 4 \sqrt{z_x  r}, \;\;\;\;  
L^2 = 4 \sqrt{z_x / r} .          \label{ex10p}
\end{equation}
Derivatives $f_x, \tilde{f}_x$ in $Q^4$ can be eliminated by adding
total $x$-derivatives 
\[D_x \left(r^{-5/2} z_x^{\;\,-3/2}2z_xr(fr-\tilde{f})\Delta_4 \right)\]
to the left hand site of the identity (\ref{ex10m}) and to $D_x P^x$.
The modified form of the identity (\ref{ex10m}) is
\begin{eqnarray*}
0 & =  & z_x^{-3/2} r^{-5/2} \left(2 z_x r (f r - \tilde{f}) D_x\Delta_4
         - 2 r_x z_x (f r - \tilde{f}) \Delta_4
         + (f r + \tilde{f}) \Delta_3 \right)  \\
& = & \;\;\; D_t \left(4\sqrt{z_x/r}(r f - \tilde{f})\right) \\
&   & +D_x \left(2 z_x^{\;\,-1/2} r^{-3/2}( - 2 f_x z_x r^2 - 
          2 \tilde{f}_x z_x r + r_x z_x  f r - r_x z_x  \tilde{f}  \right.  \\
&   &  \hspace{100pt}  \left. + z_{xx} f r^2 +    z_{xx} \tilde{f} r + 4 z_x  
         a  \tilde{f} r^2 + 4    z_x   b f r )\right)\\
&   & + L^1 C_1 + L^2 C_2.
\end{eqnarray*}
Partial integration of $L^1 C_1 + L^2 C_2$ until $f, \tilde{f}$ appear
purely algebraically makes necessarily $P^t = P^x = 0$. Because $f,
\tilde{f}$ are free we obtain the identities
\begin{eqnarray}
0 \; = \; r^{-3/2} z_x^{\;\,-3/2}
\left(\Delta_3+2 r z_x D_x\Delta_4-2r_x z_x\Delta_4\right)
& = & \;\;\,L^1_t + L^1_{xx} + c L^1 + 2 b L^2_x  \label{ex10r} \vspace{6pt} \\
0 \; = \; r^{-5/2} z_x^{\;\,-3/2}
\left(\Delta_3-2 r z_x D_x\Delta_4+2r_x z_x\Delta_4\right)
& = & - L^2_t + L^2_{xx} + c L^2 + 2 a L^1_x \label{ex10s} 
\end{eqnarray}
completing the linearization. For any solution $L^1, L^2$ of
(\ref{ex10r}), (\ref{ex10s}), equations (\ref{ex10p})
provide $r, z_x$. With $z_t$ from (\ref{ex10f}) we get $z$ as a line
integral, $w$ from (\ref{ex10wz}) and $u,v$ from $r$ and equation (\ref{ex10c}).

In the following section the effect of our method on PDEs
is investigated that linearize to inhomogeneous equations.

\section{Inhomogeneous linear DEs} \label{inhomsec} 
If the general solution of conservation law determining equations
involves a number of free constants or free functions then individual
conservation laws are obtained by setting all but one to zero.
The remaining terms are homogeneous in the surviving constant or
function. The question arises whether our conservation law based
method is suitable to find linearizations that lead to linear but
inhomogeneous equations.

{\em Example 4:} For the (ad hoc constructed) equation
\begin{equation}
0 = \Delta := 2uu_t+2uu_{xx}+2u_x^{\;\,2} + 1    \label{tri11}
\end{equation}
the conservation law identity
\begin{eqnarray*}
   (f_x+\tilde{f}_t) \Delta & = & 
D_t \left( f_xu^2+\tilde{f}_tu^2+\tilde{f} \right) +  \\
 & &  D_x \left(-f_{xx}u^2+2f_xuu_x-
                \tilde{f}_{t,x}u^2+2\tilde{f}_tuu_x+f\right) + \\
 & &  u^2(f_{t,x}-f_{xxx}-\tilde{f}_{txx}+\tilde{f}_{tt})
\end{eqnarray*}
involves functions $f(x,t), \tilde{f}(x,t)$ and establishes a conservation law
provided $f, \tilde{f}$ satisfy
\begin{equation} 
0 = f_{t,x}-f_{xxx}-\tilde{f}_{txx}+\tilde{f}_{tt}. \nonumber
\end{equation} 
Our method gives the linear system
\begin{eqnarray}
0 \;\; = \;\; D_x \Delta & = & L_{tx} + L_{xxx}   \label{tri13} \\
0 \;\; = \;\; D_t \Delta & = & L_{tt} + L_{txx}.  \label{tri14} \\
L & = & u^2  \nonumber  
\end{eqnarray}
The system (\ref{tri13}), (\ref{tri14}) represents the $x$ and $t$ 
derivatives of the linear equation
\begin{equation}
0 = L_t + L_{xx} + 1    \label{tri15}
\end{equation}
which is equivalent to equation (\ref{tri11}) and is an inhomogeneous
linear PDE.
Although our linearization method does not quite reach (\ref{tri15}),
it nevertheless
provides $L = u^2$ as the new unknown function which makes it easy to get
to the equivalent linear equation (\ref{tri15}) through a change of
dependent variables in (\ref{tri11}) or through an integration
of (\ref{tri13}), (\ref{tri14}).

The way how homogeneous consequences can be derived from
an inhomogeneous relation is to divide the inhomogeneous relation through the
inhomogeneity, i.e.\ to make the inhomogeneity equal 1 and 
then to differentiate with
respect to all independent variables and to obtain a set of linear
homogeneous conditions in the same way as equations 
(\ref{tri13}), (\ref{tri14}) are consequences of (\ref{tri15}).
The application in the following section leads to an inhomogeneous linear
PDE with non-constant inhomogeneity.

\section{An example of a triangular linear system} \label{triagsection}
A generalization of complete linearizability of the whole PDE system
in one step is the successive linearization of one equation at a time. 

{\em Example 5:} Assume a triangular system of equations, 
like the (ad hoc constructed) system
\begin{eqnarray}
 0 \;\, = \;\, \Delta_1 & := & u_t   \label{aex1} \\
 0 \;\, = \;\, \Delta_2 & := & vv_t - uvv_{xx} - uv_x^{\;\,2}  \label{aex2}
\end{eqnarray}
with one equation (\ref{aex1}) involving only one function, say
$u=u(x,t)$, and this equation being linear or being linearizable and a
second nonlinear equation being linear or linearizable in another
function $v=v(x,t)$.  How can the method in section \ref{method} be
used to recognize that such a system can be solved by solving
successively only linear equations?

In determining all conservation laws for this system with unknown
functions $v,u$ and with integrating factors of order
zero we get apart from two individual conservation laws
with pairs of integrating factors 
$(Q^1,Q^2) = (\frac{v^2}{u^2}, - \frac{2}{u})$
and $(\frac{xv^2}{u^2}, - \frac{2x}{u})$
only one with a free function $f(u,x)$:
\[ f_u \Delta_1 =  D_t f\]
which indicates the linearity of $\Delta_1$ but not the linearity of
$\Delta_2$ in $v$ once $u(x,t)$ is known.

The proper way of applying the method of section \ref{method} is to
compute conservation laws of $0=\Delta_2$ alone which
now is regarded as an equation for $v(x,t)$ only.
The function $u(x,t)$ is assumed to be parametric and given.
We obtain the identity
\[ 2f \Delta_2 = D_x (f_xuv^2 + fu_xv^2 - 2fuvv_x) +
                 D_t (fv^2) - 
     v^2(f_t + u f_{xx} + 2u_x f_x + u_{xx} f). \]
which is a conservation law
if $f$ satisfies the linear condition
\begin{equation}
 0 = f_t + u f_{xx} + 2u_x f_x + u_{xx} f.         \label{fcond}
\end{equation}
This provides the linearization
\begin{eqnarray*}
 0 \;\, = \;\, 2 \Delta_2 & = & L_t - u L_{xx} \\
 L & = & v^2 .
\end{eqnarray*}
The reason that now a linearization of $\Delta_2$ is reached is that
in the second try
$u$ is assumed to be known and therefore $u, u_{xx}, \ldots$ are not
jet-variables and hence the condition (\ref{fcond}) has solutions,
otherwise not.

Examples where this triangular linearization method is successful are the
systems (17), (18) in \cite{SW}. We demonstrate the method with
one of these (system (17)), the other is similar. 

{\em Example 6:} The system
\begin{eqnarray}
0 = \Delta_1 & := & u_t - u_{xx}-4uvu_x-4u^2v_x-3vv_x-2u^3v^2-uv^3-au 
    \label{tri1}\\
0 = \Delta_2 & := & v_t + v_{xx}+2v^2u_x+2uvv_x+2u^2v^3+v^4+av
    \label{tri2}
\end{eqnarray}
involves functions $u(x,t), v(x,t)$ and the constant $a$.
The single conservation law
\[ 0 = v \Delta_1 + u \Delta_2  = D_t (uv) + D_x (uv_x-u_xv-u^2v^2-v^3) \]
motivates the introduction of a function $w(x,t)$ through
\begin{eqnarray}
  w_x & = & uv     \label{tri3}  \\
- w_t & = & uv_x-u_xv-u^2v^2-v^3.       \label{tri4}
\end{eqnarray}
Substitution of $u$ from (\ref{tri3}) brings 
equations (\ref{tri2}) and (\ref{tri4})
in the form
\begin{eqnarray}
0 = \Delta_3 & := &w_t - \frac{1}{v}( - 2v_xw_x + w_{xx}v + w_x^{\;\,2}v + v^4)\label{tri5} \\
0 = \Delta_4 & := &v_t + v_{xx} + 2w_{xx}v + 2w_x^{\;\,2}v + av + v^4.   \label{tri6}
\end{eqnarray}
This system obeys conservation laws that involve a function $f(x,t)$ that
has to satisfy $f_t=f_{xx}+af$. Our procedure provides the linearization
\begin{eqnarray}
e^w(v\Delta_3 + \Delta_4) & = & L^1_t  + L^1_{xx}  + aL^1\;\,=\;\,0 \label{tri8} \\
L^1 & := & ve^w .  \nonumber 
\end{eqnarray}
The second linearized equation can be obtained by
\begin{itemize}
\item substituting $v=L^1/e^w$ into equations (\ref{tri5}) and (\ref{tri6}):
to get the remaining condition
\begin{equation}
0 = \Delta_5 := w_t - w_{xx} - 3w_x^{\;\,2} + 2 w_x L^1_x (L^1)^{-1} - (L^1)^3e^{-3w},
\label{tri9}
\end{equation}
\item assuming $L^1$ has been solved from (\ref{tri8}) and
treating $L^1(x,t)$ as a parametric function when
computing conservation laws for equation (\ref{tri9})
which turn out to involve two
functions that have to satisfy linear PDEs,
\item performing the linearization method to find that
the remaining equation (\ref{tri9}) linearizes with $L^2=e^{3w}$ to
\begin{equation}
e^{3w}\Delta_5 = L^2_t -  L^2_{xx}  + 2L^2_xL^1_x/L^1 - 3(L^1)^3.  
\label{tri10}
\end{equation}
\end{itemize}
Because the condition (\ref{tri10}) is inhomogeneous for $L^2$ due to
the term $3(L^1)^3$, actually two homogeneous
linear equations are generated which are
the $x-$ and $t-$ derivative of (\ref{tri10}) divided by $3(L^1)^3$ 
(see previous section about linear inhomogeneous equations). 
But as the function $L^2=e^{3w}$ results in this process, it is no
problem to find (\ref{tri10}) from (\ref{tri9}) directly
or from an integration of these two equations.
This completes the linear triangularisation of the original problem
(\ref{tri1}),(\ref{tri2}) to the new system (\ref{tri8}),(\ref{tri10}).

\section{Summary} 
The paper starts with introducing conservation law identities as a
natural way to formulate infinite parameter conservation laws.

Conservation law identities are the input to a four step procedure
that returns a differential consequence of the original system 
together with a linear differential operator that can be factored out. 

Sufficient conditions on the conservation law identity which either
guarantee a complete linearization or at least a partial linearization
are discussed.

The possibility to find a non-local linearization arises from the
application of single (finite parameter) conservation laws with the
aim to introduce potentials which satisfy infinite parameter
conservation laws and thus allow a linearization.

In examples it is demonstrated how the standard procedure 
can lead to inhomogeneous linear PDEs and 
how a successive linearization of one equation at a time may be
possible when the whole system can not be linearized at once.

\section*{Appendix} 
In the appendix we list further examples of linearizations and
integrations without giving details of the calculations. 

The first example of the Kadomtsev-Petviasvili equation 
demonstrates what our method gives when a PDE
has $p$ independent variables and the conservation law involves free
functions of less than $p-1$ variables. Although the result will be
less useful than in the other examples, we still include it
for illustration.

{\em Example 7:} 
The Kadomtsev-Petviasvili equation
\[ 
0 = \Delta = u_{tx} + u_{xxxx} + 2u_{xx}u + 2u_x^{\;\,2} - u_{yy}
\] 
for $u(t,x,y)$ has
four conservation laws with a zeroth order integrating factor
and an arbitrary function $f(t)$ as given in \cite{CL4}.
We comment on one of these four with an integrating factor
$f_ty^3+6fxy$ as the situation for the others is similar. Omitting the
details we only give the result of our method: 
\begin{eqnarray*}                      
L^1 & = & y\left(u_{txxx}y^2 + 2u_{tx}uy^2 + u_{tt}y^2 + 2u_tu_xy^2 \right. \\
    &   &  \left.\;\; - 6u_tx - 6u_{xxx}x + 6u_{xx} - 12u_xux + 6u^2 \right) \\
L^2 & = & - u_{ty}y^3  + 3u_ty^2  + 6u_yxy - 6ux   \\
y(6x\Delta-y^2D_t\Delta) & = & L^1_x + L^2_y.
\end{eqnarray*}
The arbitrary function $f(t)$ involves only one independent
variable $t$ and the conservation law $0 = L^1_x + L^2_y$ 
involves two functions $L^1,L^2$ and has
derivatives with respect to two variables $x,y$ and is therefore not
as useful as if it would be a single total derivative.

The three following equations were shown to the author 
first by V. Sokolov \cite{ZS}
who obtained their integrations earlier and independently. We
add them here to demonstrate that these results can be obtained in a
straight forward procedure.\vspace{6pt}

{\em Example 8:} For the equation
\begin{equation}
0 = \Delta := u_{xy} - e^u \sqrt{u_x^{\;\,2}-4},\;\;\;\; 
u = u(x,y) \label{ex21}
\end{equation}
a conservation law with an arbitrary function $f(x)$ enables to factor
out $D_y$ leaving an ODE to solve
\begin{equation}
(u_x-D_x)\left(\frac{\Delta}{\sqrt{u_x^{\;\,2}-4}}\right) = 
D_y\left(\frac{-u_{xx}+u_x^{\;\,2}-4}{\sqrt{u_x^{\;\,2}-4}}\right)\;\;=\;\;0.
\end{equation}
Another conservation law with an arbitrary function $g(y)$ gives
\begin{equation}
\left(u_y-\frac{u_xe^u}{\sqrt{u_x^{\;\,2}-4}}-D_y\right)\Delta =
D_x\left(-u_{yy}+\frac{1}{2}u_y^{\;\,2}+\frac{1}{2}e^{2u}\right)\;\;=\;\;0.
\label{ex22}
\end{equation}

{\em Example 9:} For the equation
\begin{equation}
0 = \Delta := u_{xy} - \left(\frac{1}{u-x}+\frac{1}{u-y}\right) u_x u_y,
\;\;\;\; u = u(x,y) \label{ex31}
\end{equation}
a conservation law with an arbitrary function $f(x)$ similarly to the
above example provides
\begin{equation}
\frac{y - x}{(u - x)(u - y)}\Delta + D_x\left(\frac{\Delta}{u_x}\right) =
D_y\left(\frac{u_{xx}}{u_x}-\frac{2(u_x-1)}{u-x}-\frac{u}{(u-x)x}\right)
\;\;=\;\;0.  \label{ex32}
\end{equation}
A second conservation law is obtained from an arbitrary function
$g(y)$ and is equivalent to (\ref{ex32}) after swapping
$x \leftrightarrow y$.\vspace{6pt}

{\em Example 10:} For the equation
\begin{equation}
0 = \Delta := u_{xy} - \frac{2}{x+y}\sqrt{u_x u_y},
\;\;\;\; u = u(x,y) \label{ex41}
\end{equation}
a conservation law with an arbitrary function $f(x)$ gives
\begin{equation}
\frac{1}{(x+y)}\left(\frac{1}{\sqrt{u_x}}-\frac{1}{\sqrt{u_y}}\right) \Delta 
+ D_x\left(\frac{\Delta}{\sqrt{u_x}}\right) =
D_y\left(  \frac{u_{xx}}{\sqrt{u_x}} + \frac{2\sqrt{u_x}}{x+y}
\right)\;\;=\;\;0. \label{ex42}
\end{equation}
A second conservation law is obtained from an arbitrary function
$g(y)$ and is equivalent to (\ref{ex42}) after swapping
$x \leftrightarrow y$.\vspace{6pt}

The final example shows a linearization of a system that
resulted in classifying non-linear 
Schr\"{o}dinger type systems in \cite{SW}.\vspace{6pt}

{\em Example 11:} 
The system
\begin{eqnarray}
0 = \Delta_1 & := & u_t  - u_{xx} - 2vu_x - 2uv_x - 2uv^2 - u^2 - au - bv - c
    \label{vol41}\\
0 = \Delta_2 & := & v_t + v_{xx} + 2vv_x + u_x
    \label{vol42}
\end{eqnarray}
involves functions $u(x,t), v(x,t)$ and the constants $a,b,c$.
The trivial conservation law
\[ 0 = \Delta_2 = D_t (v) + D_x (v_x+u+v^2) \]
motivates the introduction of a function $w(x,t)$ through
\begin{eqnarray}
  w_x & = & v                          \label{vol43}  \\
- w_t & = & v_x+u+v^2.       \label{vol44}
\end{eqnarray}
Substitution of $u,v$ from (\ref{vol43}) and (\ref{vol44})
brings equation (\ref{vol41}) in the form
\begin{eqnarray*}
0 \;\; = \;\;\Delta_3 & = &   w_{tt} + w_t^2  - w_ta - w_{xxxx}
               - 4w_{xxx}w_x - 3w_{xx}^2 - 6w_{xx}w_x^2 - w_{xx}a \\
         &   & - w_x^4 - w_x^2a + w_xb + c.
\end{eqnarray*}
This equation admits a conservation law identity
\begin{eqnarray*}
fe^w\Delta_3 & = & D_t\left[ (e^w)_tf - e^wf_t - e^wfa \right] +  \\
               & & D_x\left[ - (e^w)_{xxx}f + (e^w)_{xx}f_x - (e^w)_xf_{xx}
                     + e^wf_{xxx} \right. \\
             & & \;\;\;\;\;\, \left. - (ae^w)_xf + ae^wf_x + be^wf \right]  \\
               & & + e^w\left[f_{tt}+af_t-f_{xxxx}-af_{xx}-bf_x+cf\right].
\end{eqnarray*} From this follows the linearization
\begin{eqnarray*}
e^w\Delta_3 & = & L_{tt}-L_ta-L_{xxxx}-L_{xx}a+L_xb+Lc \;\; = \;\;0 \\
          L & = & e^w .
\end{eqnarray*}

\end{document}